\documentclass[doublecol]{epl2}
\usepackage[version=3]{mhchem} % Formula subscripts using \ce{}
\usepackage{amsmath}
\usepackage{graphicx}
\usepackage{color}
\begin{document}

\title{The interaction of charged nanoparticles at interfaces}

\author{Andreas S. Poulos\thanks{Current address: IESL, FORTH, Heraklion, Greece.} \and Doru Constantin\thanks{Corresponding author. Email: doru.constantin@u-psud.fr} \and Patrick Davidson \and Marianne Imp\'{e}ror-Clerc \and Brigitte Pansu \and St\'{e}phan Rouzi\`{e}re}
\shortauthor{Andreas S. Poulos \etal}
\institute{Laboratoire de Physique des Solides, Universit\'{e} Paris-Sud, CNRS, UMR 8502, 91405 Orsay, France.}


\abstract{
We study charged nanoparticles adsorbed onto surfactant bilayers using small-angle scattering of synchrotron radiation. The in-plane interaction of the particles is well described by a DLVO component (measured independently in solution) and a repulsive dipolar interaction due to the presence of the interface, with an amplitude close to the theoretical prediction. We prove that charged nanoparticles at soft interfaces are well described by the classical model of Poisson-Boltzmann and van der Waals terms; as a corollary, they do not experience the like-charge attraction reported in the literature for some systems of micron-sized spheres at interfaces.
}

\maketitle

\section{Introduction}\label{intro}
The distribution of ions close to soft interfaces (such as biological membranes) is a long-standing problem \cite{Berkowitz:2006}, with wide-ranging fundamental and practical implications. The variation in ion density as a function of the distance to the interface has been thoroughly studied, mostly by X-ray scattering techniques \cite{Vaknin:2003,Luo:2006,Park:2006}, but very little is known about the individual ion-ion interaction, although this interaction and the resulting ion correlation play an important role in biological processes \cite{Grosberg:2002}. Such information could also help validate and refine certain assumptions made in theoretical and numerical studies of charged systems at the nanoscale, e.g.\ in implicit solvent strategies \cite{Feig:2004,Baker:2005}.

One often assumes that the interaction is accurately described by the DLVO model \cite{Israelachvili:2011}, which combines the electrostatic and van der Waals interactions, supplemented by steric repulsion. While this model has been very successful in explaining the interaction of macroscopic objects, its validity at the nanoscale is still in need of direct experimental verification.

\begin{figure}[htbp]
\centerline{\includegraphics[width=0.5\textwidth]{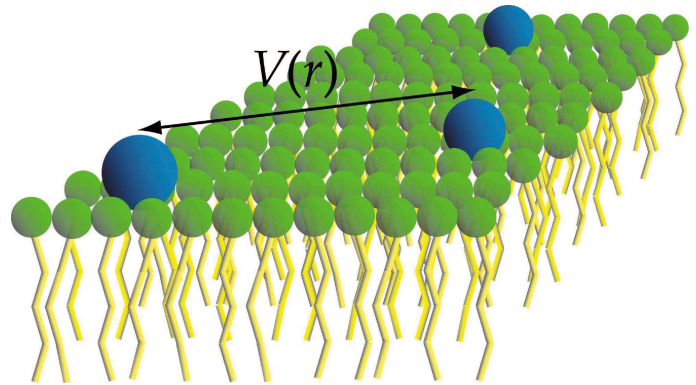}}
\caption{Schematic representation of charged nanoparticles adsorbed onto a surfactant bilayer (only the top monolayer is represented).}
\label{fig:illustr}
\end{figure}

In this context, measuring the correlations between highly charged nanoparticles adsorbed at an interface could provide edifying results. However, due to their small size and often low scattering contrast, very little data exists for such systems. In the present study, we use small (1~nm in diameter) and multivalent inorganic anions adsorbed onto bilayers formed by a nonionic surfactant (Figure \ref{fig:illustr}) and determine their interaction in the plane of the layers.

Due to the presence of heavy atoms, the structure factor of these particles can be measured (even at low concentration and in a disordered state) using the intense X-ray beams produced by third-generation synchrotron facilities. Using classical concepts from liquid state theory we then describe the structure factor in terms of the interaction potential and determine the relevant parameters of the latter. 

\section{Materials and methods}

\subsection{Preparation}
The system under investigation is the nonionic lyotropic lamellar phase of the surfactant Brij~30, doped with polyoxometalates (POMs): the phosphotungstate anions $[\ce{PW_{12}O_{40}}]^{3-}$. In previous work \cite{Poulos:2008,Poulos:2010} we have shown that the POMs are strongly associated to the heads of the surfactant molecules, forming one layer on either side of the surfactant membrane. The preparation parameters are the overall volume fraction of surfactant $\phi_S$ and the volume fraction of POMs in the aqueous medium, $\phi_P$. Perfectly oriented homeotropic monodomains are prepared inside flat optical capillaries by thermal annealing. No salt was added to the samples, so the only available counterions are the protons resulting from the (complete) dissociation of the phosphotungstic acid. Their water concentration varies between 73 and 436~mM, so we can safely neglect the possible contaminating ions.

All chemicals were purchased from Sigma-Aldrich and used without any further purification. Brij~30 consists mostly of \ce{C_{12}EO_4} (tetraethylene glycol dodecyl ether), along with a smaller amount of homologous \ce{C_{\textit{m}}EO_{\textit{n}}} molecules. Its phase diagram is very similar to that of pure \ce{C_{12}EO_4} \cite{Strey:1996}. At room temperature, it forms a lamellar phase over a wide concentration range, between 25 and 85~wt\%. The surfactant bilayer is 34~{\AA} thick \cite{Poulos:2008}.

The phosphotungstic acid is completely dissociated in water, giving \ce{[PW_{12}O_{40}]^{3-}} polyoxometalates (POMs) and \ce{H^+}. The mass fraction of hydration water in the initial powder was determined (by drying at 200~$^\circ$C) as approximately 15\%, in good agreement with the 20\% value reported in the literature \cite{Marosi:2000} and was accounted for in the preparation. Aqueous POM solutions at a volume fraction $\phi_P$ were prepared by dissolving a known quantity of powder in ultrapure water, as discussed in detail in a previous publication \cite{Poulos:2010}. These solutions were then mixed with appropriate amounts of surfactant to yield the desired surfactant volume fraction $\phi_S$.

\subsection{Sample alignment}
The POM-doped lamellar phase was drawn (by suction with a syringe) into flat optical glass capillaries, 0.1~mm thick and 2~mm wide (VitroCom Inc., Mountain Lakes, NJ) which were then flame-sealed. Using a Mettler FP52 heating stage, the capillaries were heated above the lamellar-to-isotropic transition temperature and cooled slowly (0.1~$^\circ$C/min) down to room temperature. Upon nucleation, the lamellar phase domains are aligned with their bilayers parallel to the flat faces of the capillary (homeotropic anchoring). These domains grow to span the whole width of the capillary \cite{Poulos:2008}.

\subsection{SAXS experiments}
All data shown in this paper was recorded at the ID02 station of the European Synchrotron Radiation Facility (ESRF, Grenoble, France), with an incident beam energy of 12.4~keV and at a sample--detector distance of 0.885~m. The scattered x-rays were detected with a specially developed CCD camera \cite{Narayanan:2001}. The flat faces of the capillaries were set perpendicular to the x-ray beam (normal incidence). The radially averaged intensity $I(q)$ is obtained in absolute units; it is subsequently divided by the form factor $|Ff(q)|^2$  of a POM (calculated from the coordinates of the constituent atoms) and normalized by the known density of particles to yield the structure factor $S(q)$ over a scattering vector range $0.02 < q < 0.5 \, \text{\AA}^{-1}$.

Measurements under oblique incidence were done at the bending magnet beamline BM02 (D2AM) \cite{Simon:1997} of the ESRF, at a photon energy of 11~keV and a sample--detector distance of 0.270~m. The detector was a CCD Peltier-cooled camera (SCX90-1300, from Princeton Instruments Inc., New Jersey, USA) with a detector size of $1340 \times 1300$ pixels. The data was pre-processed using the \texttt{bm2img} software developed at the beamline. Further treatment followed closely a method presented in a previous publication \cite{Constantin:2010b} and is described in the Supplementary Material.

\subsection{Analysis}
For a given interaction potential, the theoretical structure factor is calculated iteratively (based on the Ornstein-Zernicke equation with the Percus-Yevick closure) using the method of Lado \cite{Lado:1967,Lado:1968}; details on the implementation are given in references \cite{Constantin:2009,Constantin:2010b}. With the appropriate modifications, this procedure was used for modelling the (three-dimensional) interaction of POMs in solution as well as for the (two-dimensional) interaction of POMs confined within the lamellar phase, as discussed below.

\section{Results}\label{results}

\subsection{Interaction in aqueous solution}\label{solution}
\begin{figure}[htbp]
\centerline{\includegraphics[width=0.45\textwidth]{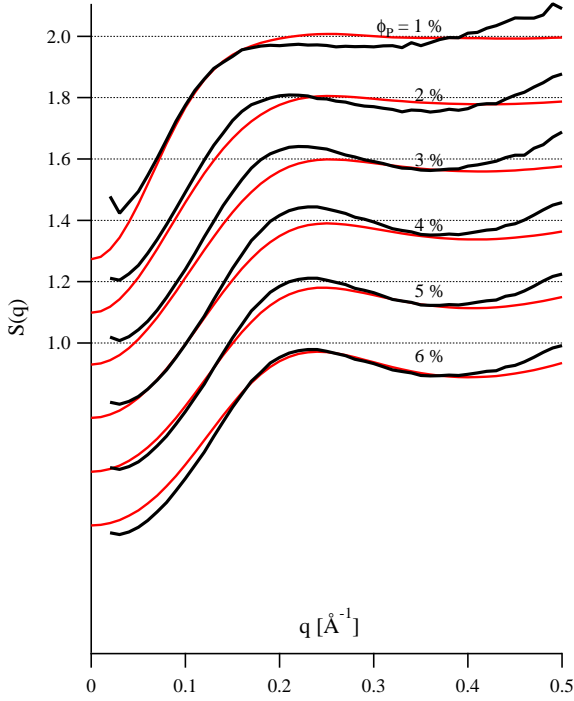}}
\caption{Structure factors of POMs in aqueous solution, for different volume fractions $\phi_P$ indicated alongside the curves (black), as well as fits using the DLVO potential (\ref{eq:DLVO}), in red; the curves are shifted vertically in steps of 0.2, starting from the most concentrated at the bottom. The parameter values are indicated in the text.}
\label{fig:intersol}
\end{figure}

In order to evaluate the effect of the interfacial confinement on the interaction between POMs one needs an accurate determination of their interaction in solution, in the absence of confinement. We measured the structure factors $S(q)$ for aqueous solutions of POMs, with a $\phi_P$ between 1 and 6~vol\% (Fig.~\ref{fig:intersol}). In the framework of the DLVO theory, the interaction outside the hard core consists of an electrostatic repulsion and a van der Waals attraction, considered additive. The resulting potential is:
\begin{equation}
\begin{array}{l l}
V_{\text{sol}}(r) &=
\left \lbrace
  \begin{array}{l l}
    &\infty \quad \text{for} \,\, r < 2 R_C\\
    & V_{\text{el}}(r) + V_{\text{vdW}}(r)\quad \text{for} \,\, r \geq 2 R_C\\
  \end{array}
\right.\\
& \\
    V_{\text{el}}(r) &= k_B T \, Z_{\text{eff}}^2 \left [ \frac{\exp (\kappa R_C)}{1+ \kappa R_C} \right ]^2 \frac {\exp (-\kappa r)}{r/\ell_{B}}\\
   V_{\text{vdW}}(r) &= - \frac{A}{6} \left [\frac{1}{\rho - 2} + \frac{1}{\rho} + \ln \left (1 - \frac{2}{\rho} \right ) \right ]
\end{array}
\label{eq:DLVO}
\end{equation}
\noindent where $Z_{\text{eff}} = \alpha Z_{\text{bare}}$ is the effective charge; $Z_{\text{bare}} = 3$ is the bare charge of a POM and the renormalization parameter $\alpha = 0.9$ \cite{Aubouy:2003}. $\ell_{B} = 7 \, \text{\AA}$ is the Bjerrum length for water, $\kappa$ is the reciprocal Debye length (and it depends implicitly on the POM concentration). In the van der Waals term (written for solid spheres) we defined $\rho = (r/R_0)^2/2$, with $R_0 = 5.2 \, \text{\AA}$ the crystallographic radius of the POMs (position of the outermost oxygen atom).
The only adjustable parameters in the model are the effective core radius $R_C$ and the Hamaker constant $A$. The fit yields $R_C = 5.53 \, \text{\AA}$ and $A = 13 \, k_B T$, in reasonable agreement with the literature values for metal oxides: $A = 5-12.5 \, k_B T$ \cite{Bergstrom:1997}.

\subsection{No interaction along the bilayer normal}\label{zinter}
For particles dispersed into the lamellar phase, one must consider the interaction both in-plane and along the layer normal, and hence the dependence of the structure factor on both $q_r$ and $q_z$. Most of the experiments discussed here were performed with the incident beam parallel to $\hat{z}$ (i.e. perpendicular to the bilayers and to the flat faces of the capillary), so that the scattering vector is contained within the plane of the layers: $q = q_r$ (we neglect the curvature of the Ewald sphere). The measured $S(q)$ could nevertheless be influenced by a possible interaction along $z$ \cite{Yang:1999}. In particular, should such an interaction exist, $S(q)$ would vary significantly with the confinement imposed by $\phi_S$, which imposes the distance between two particle layers facing each other across a water space.

\begin{figure}[htbp]
\centerline{\includegraphics[width=0.45\textwidth]{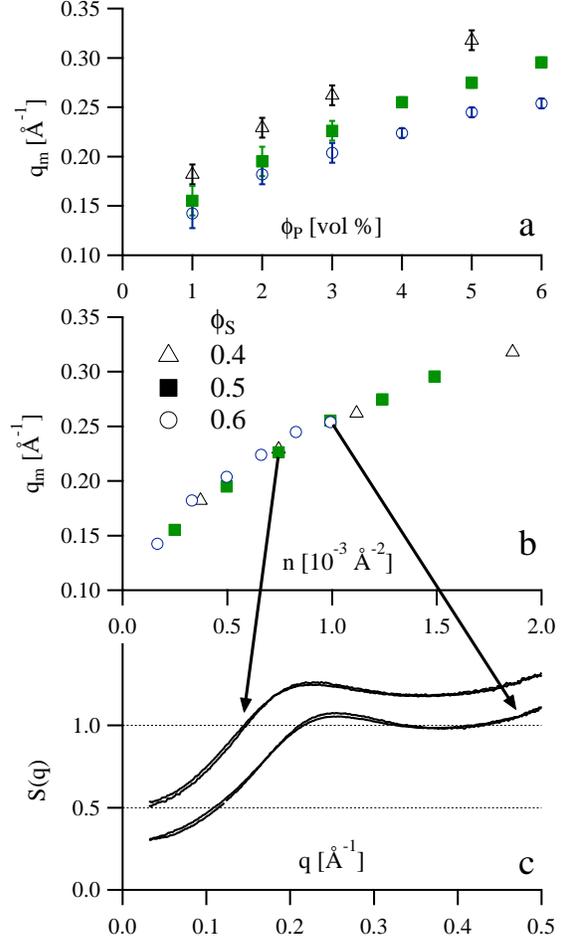}}
\caption{Position of the structure factor maximum $q_m$, plotted as a function of the volume fraction of particles $\phi_P$ (a) and as a function of their in-plane density $n$ (b) for different surfactant concentrations $\phi_S$ (various symbols and colors). In (c) we compare structure factors with the same $n$, but different $\phi_P$ and $\phi_S$. The two curves corresponding to $n= 0.75 \, 10^{-3}\, \text{\AA}^{-2}$, with $\phi_S=0.4$ and 0.5, are shifted upwards by 0.2.}
\label{fig:nointer}
\end{figure}

The position of the maximum $q_m$ of the structure factor is sensitive to the interaction. We show this parameter in Fig.~\ref{fig:nointer}a as a function of $\phi_P$ and in Fig.~\ref{fig:nointer}b as a function of the in-plane density defined as $n=\frac{\phi_P d_B}{2V_{0}}\frac{1-\phi_S}{\phi_S}$, with $d_B = 34 \, \text{\AA}$ the bilayer thickness and $V_0 = 685\, \text{\AA}^3$ the volume occupied by a POM in the crystal.
There is no simple dependence of $q_m$ with $\phi_P$ (Fig.~\ref{fig:nointer}a), but all the values fall onto the same curve when plotted against $n$ (Fig.~\ref{fig:nointer}b). Moreover, structure factors measured on samples with the same $n$ but different $\phi_P$ and $\phi_S$ share not only the maximum position, but also the details of $S(q)$, as shown in Fig.~\ref{fig:nointer}c for two sets of curves with $n= 0.75$ and $1\, 10^{-3}\, \text{\AA}^{-2}$ (the first set is shifted upwards by 0.2). We conclude that the relevant parameter for describing the interaction is the in-plane density $n$, and not the bulk density $\phi_P$; the lamellar phase can be seen as a stack of particle layers (two for each surfactant bilayer) without interaction from layer to layer.

We confirmed this conclusion by measurements where the capillary normal is tilted away from the incident beam (oblique incidence); this way, the scattering vector $\mathbf{q}$ acquires a $q_z$ component and one can determine the interaction between particles along $z$ \cite{Yang:1999,Constantin:2010b}. Such an interaction is indeed detected for $\phi_S = 0.7$, but not at $\phi_S$ up to 0.6, as shown in the Supplementary Material. In the present analysis we only include the latter data points.

\subsection{Interaction in the plane of the interface}\label{rinter}
\begin{figure}[htbp]
\centerline{\includegraphics[width=0.45\textwidth]{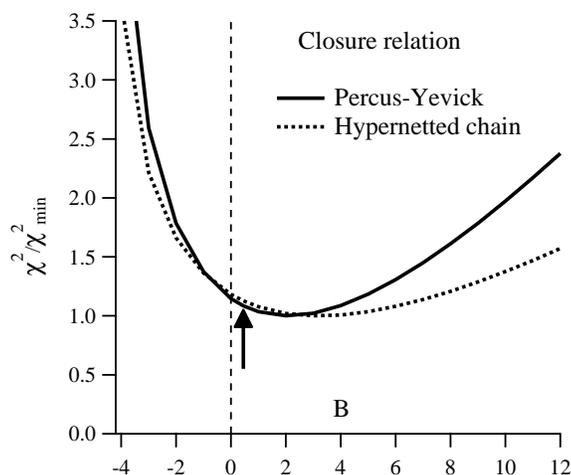}}
\caption{Goodness of fit function $\chi ^2$ normalized by its minimum value $\chi ^2 _{\text{min}}$ as a function of the parameter $B$ defined in Eq.~(\ref{eq:dipolar}), for the Percus-Yevick and hypernetted chain closures (solid and dotted line, respectively). The arrow corresponds to the theoretical prediction $B=0.45$.}
\label{fig:Chisquare}
\end{figure}

When adsorbed onto the bilayer the particles still experience the DLVO interaction, but the presence of the interface induces additional effects. In particular, due to the difference in ionic strength between the water medium and the bilayer, each particle acquires a dipole moment $P$ \cite{Pieranski:1980}, leading to a power-law repulsion \cite{Hurd:1985,Nikolaides:2002}:
\begin{equation}
V_d = 2 \frac{P^2}{4\pi \epsilon _0 \epsilon _w r^3}  = B \frac{\ell_{B}}{\kappa ^2} \frac{k_B T}{r^3}
\label{eq:dipolar}
\end{equation}
\noindent where the dielectric constants are $\epsilon _w \approx 80$ for water and $\epsilon _b \approx 2$ for the bilayer \cite{White:1978}. The magnitude of the dipole can be estimated as $P = \frac{Z}{\kappa} \sqrt{\epsilon _b / \epsilon _w}$ \cite{Hurd:1985}, yielding a prefactor $B = 2 Z^2 (\epsilon _b / \epsilon _w) \approx 0.45$. We consider that the interaction between nanoparticles adsorbed onto the bilayers is given by the DLVO term (\ref{eq:DLVO}) with the same parameter values as in the bulk, but corrected for the presence of the image charge \cite{Stillinger:1961,Hurd:1985} and supplemented by the dipolar term (\ref{eq:dipolar}):
\begin{equation}
V(r) = 2 \frac{\epsilon _w}{\epsilon _w + \epsilon _b} V_{\text{el}}(r) + V_{\text{vdW}}(r) + V_d(r)
\label{eq:total}
\end{equation}
\begin{figure}[htbp]
\centerline{\includegraphics[width=0.45\textwidth]{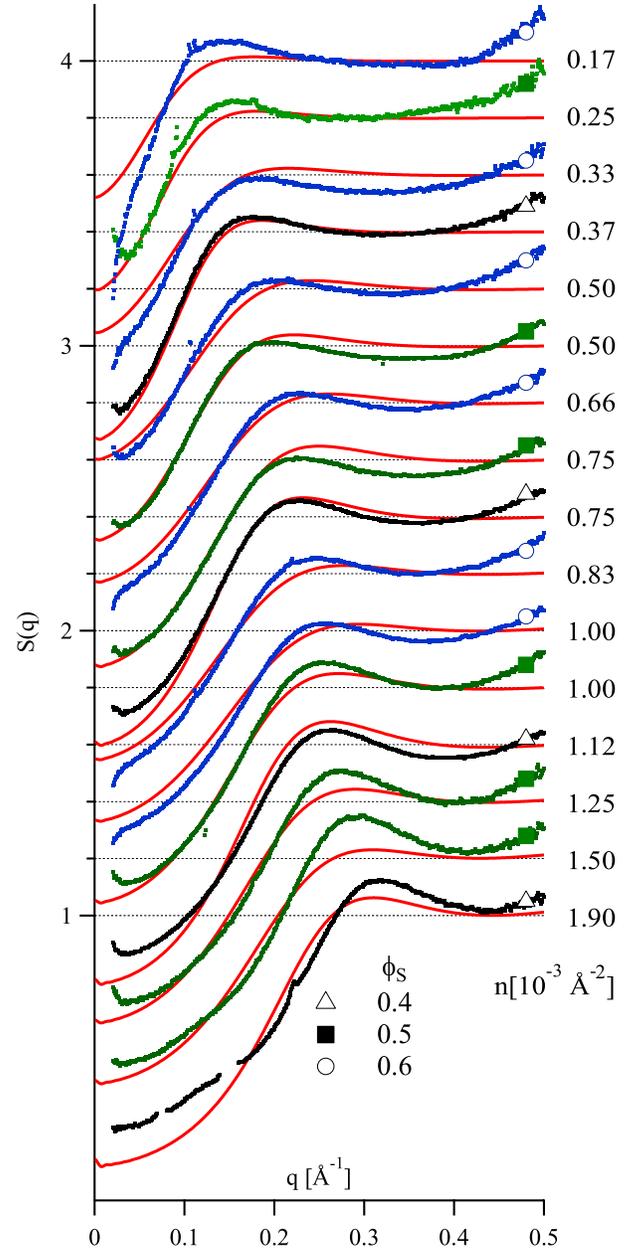}}
\caption{Structure factors of POMs adsorbed onto surfactant bilayers, for different in-plane number densities $n$ from 0.17 to $1.9 \, 10^{-3}\, \text{\AA}^{-2}$ (indicated alongside each curve), as well as best fits using the total potential (\ref{eq:total}); the curves are shifted vertically in steps of 0.2, starting from the most concentrated at the bottom. Data corresponding to different confinement values $\phi _S$ are identified by distinct symbols and colors.}
\label{fig:interplan}
\end{figure}
The experimental data are shown in Figure \ref{fig:interplan}, for surface densities $n$ from 0.17 to $1.9 \, 10^{-3}\, \text{\AA}^{-2}$. We emphasize that there is a single fit parameter --the prefactor $B$ in Eq.~(\ref{eq:dipolar})-- for all the 16 curves at various concentrations. The best fit is obtained for $B = 2$, somewhat larger than the theoretically expected value of $B = 0.45$. To check the significance of this discrepancy, we plot in Figure \ref{fig:Chisquare} the goodness-of-fit function $\chi ^2$, normalized by its minimum value $\chi ^2 _{\text{min}}$, as a function of the fit parameter $B$. It is difficult to define rigorously the uncertainty on the parameter $B$, since the difference between the data and fits in Figure \ref{fig:interplan} is systematic (due in part to imperfect background subtraction), rather than statistical. Another source of systematic error is the approximation inherent in any closure relation for the Ornstein-Zernike formula. Redoing the analysis with the hypernetted chain closure (widely employed for modelling charged systems) yields $B = 3$, with a similar fit quality.
The $\chi ^2 (B)$ function being quite flat close to the origin for both closures, the experimental result for $B$ is not significantly different from the theoretical prediction. On the other hand, the fit quality decreases rapidly for negative values of $B$, ruling out a hypothetical attractive component. Thus, we do not detect the like-charge attraction recently discussed in the case of microspheres at the air-water interface \cite{Larsen:1997,Nikolaides:2002,Aveyard:2002,Chen:2005}.

\section{Conclusion}
We infer that the two-dimensional structure factor of trivalent anions of nanometer size adsorbed onto nonionic surfactant bilayers is well described by the DLVO model and a supplementary dipolar term due to the presence of the interface. The water medium and the bilayer are treated as continuous dielectrics. This approximation is often used in the study of biomolecular systems (such as membrane proteins) \cite{Im:2003,Im:2005} but, to our knowledge, it has never before been tested experimentally for the interaction between individual ions. The data presented here support this assumption within the experimental precision.

\acknowledgments
A.~S.~P. gratefully acknowledges support from a Marie Curie action (MEST-CT-2004-514307) and from a Triangle de la Physique contract (OTP 26784). The ESRF is acknowledged for the provision of beamtime (experiments SC-2393 on the ID02 beamline and 02-01-756 on the BM02 beamline).

\bibliographystyle{eplbib}
\bibliography{POMstruct}

%\newpage
%\onecolumn
%\appendix
\section{Supplementary Material -- No interaction from layer to layer}

\subsection{Experimental}
In the main paper we analyzed the structure factors $S(q)$ measured in normal incidence and concluded that there is no interaction from layer to layer.

We confirmed this conclusion by measurements where the capillary normal is tilted away from the incident beam (oblique incidence); this way, the scattering vector $\mathbf{q}$ acquires a $q_z$ component and one can measure the complete structure factor $S(q_r,q_z)$ \cite{Yang:1999,Constantin:2010b}.

\begin{figure*}[htbp]
\centerline{\includegraphics[width=0.9\textwidth]{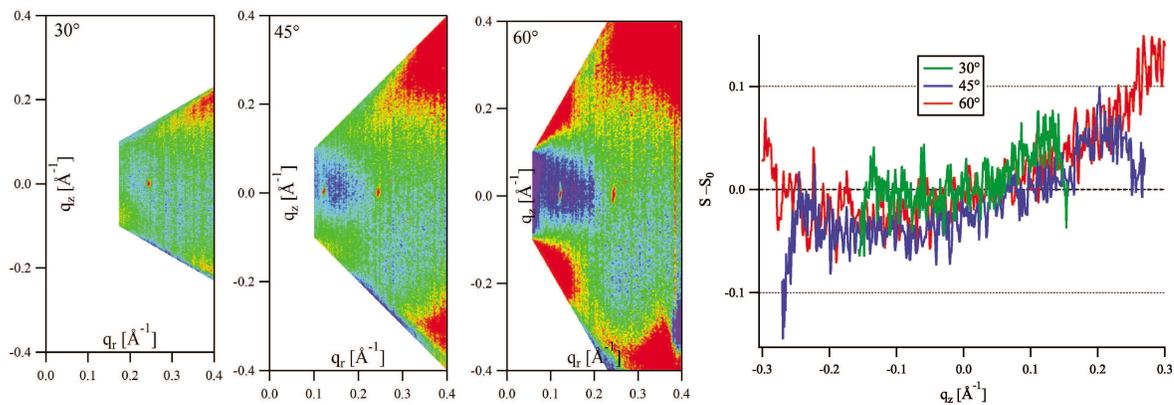}}
\caption{Complete structure factor $S(q_r,q_z)$ after subtraction of $S_0(q_r)$ for the sample with $\phi_S = 0.6$ and $\phi_P = 6\%$, for different tilt angles of the sample (left). Cuts through $S$ along $q_z$ at $q_r = 0.27\, \text{\AA}^{-1}$ (right).}
\label{fig:SM_606}
\end{figure*}

\begin{figure*}[htbp]
\centerline{\includegraphics[width=0.9\textwidth]{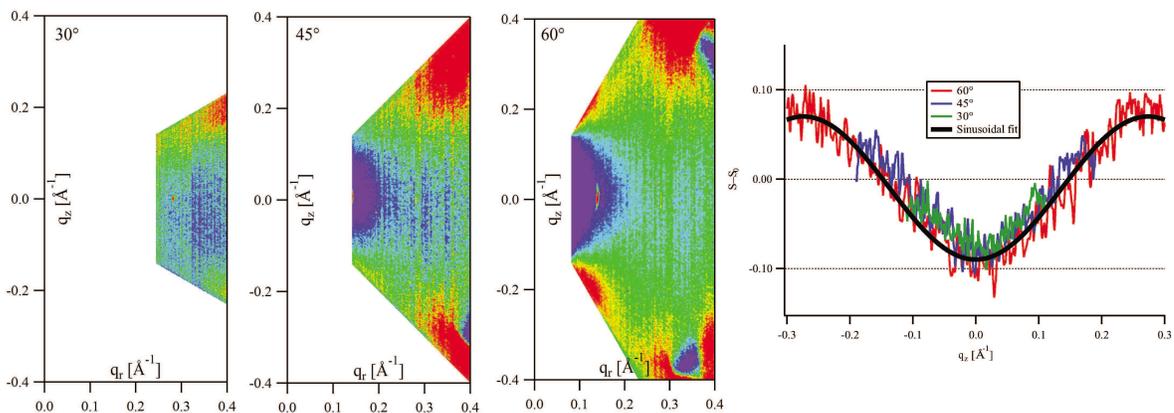}}
\caption{Complete structure factor $S(q_r,q_z)$ after subtraction of $S_0(q_r)$ for the sample with $\phi_S = 0.7$ and $\phi_P = 6\%$, for different tilt angles of the sample (left). Cuts through $S$ along $q_z$ at $q_r = 0.27\, \text{\AA}^{-1}$ and sinusoidal fit (right).}
\label{fig:SM_706}
\end{figure*}

Then, the $q_z$-independent component $S_0(q_r)$ is subtracted. The result should be sinusoidally modulated along $q_z$ if an interaction is present and essentially zero otherwise \cite{Constantin:2010b}. We present below the results for $\phi_P = 6\%$, with $\phi_S = 0.6$ and 0.7 (Figures \ref{fig:SM_606} and \ref{fig:SM_706}, respectively).

Within experimental precision, no interaction can be detected for $\phi_S = 0.6$, while at $\phi_S = 0.7$ a repulsive interaction ($S - S_0$ negative for $q_z = 0$) is clearly visible. A sinusoidal fit yields the distance between the two particle layers (facing across a water layer) as $\delta = 11.5$~\AA. For the analysis presented in the main paper we only include the data points without interaction.

\subsection{Theoretical}
Based on the above data, we can say quite confidently that there is no interlayer interaction up to $\phi_S = 0.6$; here, we check whether our theoretical model for the interaction agrees with this empirical conclusion.

\begin{figure*}[htbp]
\centerline{\includegraphics[width=0.9\textwidth]{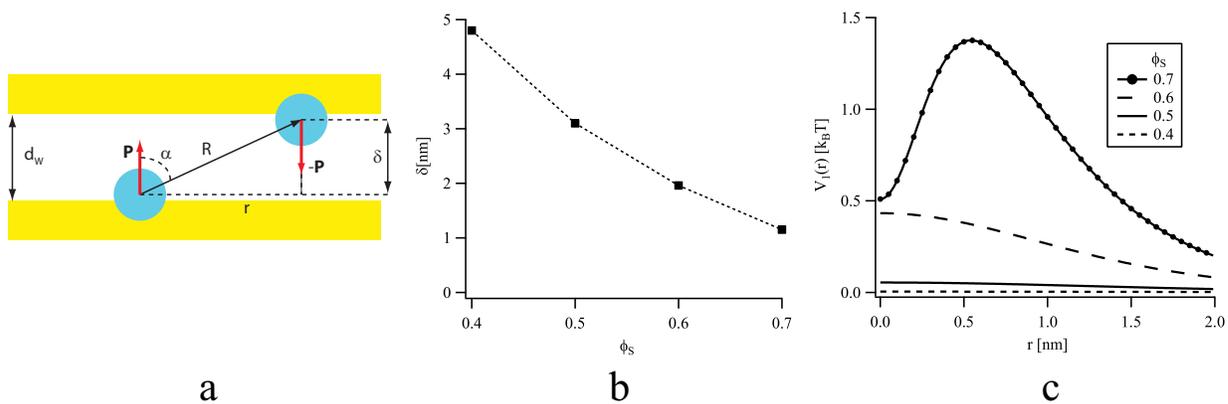}}
\caption{a) Diagram of the interaction between particles in different layers, across a water space. b) Vertical distance between the layers as a function of the surfactant volume fraction $\phi_S$. c) Interlayer interaction potential $V_1$ as a function of the in-plane distance $r$ for the highest particle concentration, $\phi_P = 6 \%$. The top two curves correspond to the experimental situation in Figures \ref{fig:SM_606} and \ref{fig:SM_706}.}
\label{fig:Interlayer}
\end{figure*}

The comparison is presented in Figure \ref{fig:Interlayer}. As shown in subfigure a), in this configuration the interacting dipoles are anti-parallel, so that the corresponding interaction is mostly attractive. The distance between the two layers $\delta$ is measured directly at $\phi_S = 0.7$ (see above) and estimated at lower surfactant concentration via the dilution law: it increases by almost a factor of two as $\phi_S$ decreases from 0.7 to 0.6 (subfigure b).

Finally, we estimate the interlayer interaction potential as a function of the in-plane distance $V_1(r)$ for all surfactant concentration at the highest particle concentration $\phi_P = 6 \%$ (Figure \ref{fig:Interlayer} c). The overall potential is positive; there is no hard-core component, since $\delta > 2 R_C$ even at the highest $\phi_S$. 
For $\phi_S = 0.4$ and 0.5, $V_1(r)$ is much lower than $k_B T$. For $\phi_S = 0.6$ it becomes of the order of $0.5\,k_B T$, but is still very slowly varying. In contrast, for $\phi_S = 0.7$ $V_1(r)$ reaches a maximum of about $1.5\,k_B T$ and varies markedly for $r < 2 \, \text{nm}$. We conclude that our model is in good agreement with the experimental findings, namely no detectable interlayer interaction for $\phi_S = 0.6$ (Figure \ref{fig:SM_606}) and mild repulsion for $\phi_S = 0.7$ (Figure \ref{fig:SM_706}).

\end{document}